\documentclass[12pt]{article}
\usepackage{graphicx}
\usepackage{amssymb}
\usepackage[nosort]{cite}

\newcommand{\bea}{\begin{eqnarray}}
\newcommand{\eea}{\end{eqnarray}}
\newcommand{\simgt}{\hbox{ \raise3pt\hbox to 0pt{$>$}\raise-3pt\hbox{$\sim$} }}
\newcommand{\simlt}{\hbox{ \raise3pt\hbox to 0pt{$<$}\raise-3pt\hbox{$\sim$} }}

\begin{document}

\begin{titlepage}
\title{\large \bf
Fine and Hyperfine Splittings of Charmonium and Bottomonium: \\
An Improved Perturbative QCD Approach
\vspace{2cm}}
\author{S.~Recksiegel$^1$ and Y.~Sumino$^2$
\\
\\ $^1$ Physik Department, Technische Universit\"at M\"unchen, \\
D--85748 Garching, Germany\\
\\ $^2$ Department of Physics, Tohoku University\\
Sendai, 980-8578 Japan
}
\date{}
\maketitle
\thispagestyle{empty}
\vspace{-5truein}
\begin{flushright}
{\bf hep-ph/0305178}\\
{\bf TU--691}\\
{\bf TUM-HEP-508/03}\\
{\bf May 2003}
\end{flushright}
\vspace{4.5truein}
\begin{abstract}
\noindent
We extend the formalism based on perturbative QCD that was developed 
in our previous work, and
compute the hyperfine splittings of the bottomonium spectrum as well as
the fine and hyperfine splittings of the charmonium spectrum.
All the corrections up to ${\cal O}(\alpha_S^5 m)$ are included in the
computations.
We find agreement (with respect to theoretical uncertainties)
with the experimental values whenever available
and give predictions for not yet observed splittings.
\end{abstract}

\end{titlepage}

Theoretical predictions of the spectra of heavy quarkonia have
traditionally been based on phenomenological models for the
inter-quark potential that are not a priori connected to 
fundamental QCD parameters \cite{Eichten:1979ms,Eichten:1994gt}. 
With suitable model potentials, very good agreement with 
the observed spectra can be obtained for the charmonium and 
bottomonium states (e.g.\ \cite{Motyka:1997di}).
These studies established the non-relativistic nature of the
heavy quarkonium systems and, in overall, a unified shape of the 
inter-quark potential in the distance region 
$0.5 \simlt r \simlt 5~{\rm GeV}^{-1}$.
More recently, 
lattice computations of the static QCD potential
gave a potential shape consistent with the phenomenological
potentials in this region \cite{bali}.
Quarkonium spectra have also been calculated in
lattice QCD \cite{Thacker:1990bm,Davies:1994mp}. 
In recent calculations of the heavy quarkonium spectra on
anisotropic lattices \cite{Chen:2000ej,LM,CP-PACS}, reasonably good 
agreement with experimental values is obtained. 
In general, lattice
calculations still suffer from uncertainties related to the
continuum extrapolation and the quenched approximation.

Recent progress in perturbative QCD has drastically
improved the description of heavy quarkonium states within perturbative QCD. 
The essential aspect is that the sum of the
QCD potential and the quark-antiquark pole masses,
$E_{\rm tot}(r) = 2m_{\rm pole}+V_{\rm QCD}(r)$, 
can be predicted
much more accurately perturbatively in the distance region relevant to 
heavy quarkonium states \cite{SNS,QCDpot,Recksiegel:2001,CplusL}, 
once the leading-order renormalons are cancelled \cite{renormalon}.
In \cite{bsv1,bsv2} the whole structure of the bottomonium spectrum
up to ${\cal O}(\alpha_S^4 m)$ was predicted within
perturbative QCD taking into account the cancellation of the leading-order
renormalons, and a good agreement with the experimental values
has been found for the gross structure of the spectrum.
Subsequently, in \cite{Recksiegel:2002za} 
we developed a specific formalism based on perturbative QCD
and examined predictions for the bottomonium spectrum.
We included in the zeroth-order Hamiltonian
the static QCD potential computed in \cite{Recksiegel:2001},
which takes into account cancellation of the renormalons.
We have included all the corrections up to 
${\cal O}(\alpha_S^5 m)$ for the fine splittings 
and all the corrections up to 
${\cal O}(\alpha_S^4 m)$ for the individual energy levels.
It was shown that good agreement between 
the computed and the observed bottomonium spectrum can be obtained,
including the fine splittings. 
These analyses \cite{bsv1,bsv2,Recksiegel:2002za} 
have shown that the predictions agree with the
corresponding experimental data within the estimated
perturbative uncertainties, and that the size of
non-perturbative contributions is compatible with the size of
perturbative uncertainties.
(Similar conclusions have been drawn in the analyses of the
perturbative QCD potential \cite{SNS,QCDpot,Recksiegel:2001,CplusL}.)

In this paper we extend our previous analysis 
\cite{Recksiegel:2002za} and predict the
hyperfine splittings of the bottomonium spectrum including 
all the corrections up to ${\cal O}(\alpha_S^5 m)$.
We also predict the fine splittings and hyperfine splittings
of the charmonium spectrum including all the corrections up to
the same order and using the same QCD potential as used for the
bottomonium spectrum.
The calculations of the bottomonium hyperfine splittings
constitute predictions of the yet unobserved states.
An essential difference of our predictions from those of the
conventional potential models is as follows:
All our predictions are constructed from the ingredients of
perturbative QCD.
Although our predictions will depend on details of 
the prescriptions adopted in our formalism, we can
identify unambiguously, within the frame of perturbative QCD, 
which higher order corrections are incorporated in our predictions, 
hence we may estimate the
sizes of the uncertainties originating from the higher-order
corrections not included in our computations.
As a result, we are able to test the validity of perturbative QCD
(within the prescriptions of our formalism), by comparing
our predictions with the experimental data with
respect to the estimated uncertainties.
This is not the case for the conventional potential models,
where their core potentials follow from simple ans\"atze.

It is not our intention to compete with treatments like \cite{Motyka:1997di}
in terms of agreement with experiment. 
Rather than constructing a very
special model potential to reproduce experimental spectra, we will
show that our predictions can successfully
describe the observed quarkonium states (and make stable
predictions for yet unobserved states), 
and that, consequently, perturbative QCD is compatible with the experimental
data with essentially
only the quark masses and the strong coupling constant as the input
parameters.

In general, where the masses of bottomonium and charmonium states
have been measured, the experimental uncertainties are much smaller
than the theoretical uncertainties \cite{Hagiwara:fs}.
The only two exceptions are the very poorly measured 
masses of $\eta_b(1S)$
and $\eta_c(2S)$. The mass of $\eta_c(2S)$ had been given as
$3594\pm 5 \,{\rm MeV}$ \cite{Edwards:1981mq} for more than 20 years,
before a new value of $3654 \pm 6\pm 8 \,{\rm MeV}$ was reported
by Belle last summer \cite{Choi:2002na}.

For the reader's convenience, let us briefly summarise the formalism
used in our previous analysis for the bottomonium spectrum 
\cite{Recksiegel:2002za}.
(The reader is referred to this paper and references therein
for technical details and an
explanation of the motivation for this choice.)
\begin{enumerate}
\renewcommand{\labelenumi}{(\arabic{enumi})}
\item 
We adopt a special organisation of the perturbative series.
The zeroth-order part of the Hamiltonian is chosen as
$H_0 = \vec{p}^{\,\,2}/(2m_{\rm pole}) + 2m_{\rm pole} + V_{\rm QCD}(r)$,
and all other operators in the Hamiltonian
are treated as perturbations.
\item 
In $H_0$, we replace
$2m_{\rm pole} + V_{\rm QCD}(r)$ by
$E_{\rm imp}(r)$, which is determined separately in three regions of $r$:
in the intermediate-distance region,
$E_{\rm imp}(r)$ is identified with $2m_{\rm pole} + V_{\rm QCD}(r)$ up to 
${\cal O}(\alpha_S^3)$, together with a specific 
prescription for fixing the renormalisation scale $\mu = \mu_2(r)$;
in the short-distance region, $E_{\rm imp}(r)$ is identified with a
three-loop renormalisation-group improved potential;
in the long-distance region a linear extrapolation is used.
\item 
As for perturbation, we include all the ${\cal O}(1/c^2)$ operators;
also some of the ${\cal O}(1/c^3)$ operators are
included, in particular, all those operators which contribute
to the ${\cal O}(\alpha_S^5m)$ corrections to the fine splittings 
are included.
\item
Our predictions depend on the renormalisation 
scale $\mu$ (for the $\overline{\rm MS}$ coupling constant), which enters
the ${\cal O}(1/c^2)$ and ${\cal O}(1/c^3)$ operators.
On the other hand, $E_{\rm imp}(r)$ in the zeroth-order Hamiltonian
is constructed such that it is independent of $\mu$, or, its
$\mu$-dependence has been removed using a specific scale-fixing
prescription.
In this way, we take into account a large difference of the 
scales involved in the 
individual energy levels and in the level splittings.
\end{enumerate}
In this work the following points are changed as compared to the above
formalism.
\begin{enumerate}
\renewcommand{\labelenumi}{(\Roman{enumi})}
\item 
Instead of the linear extrapolation, 
the long-distance part ($r>r_{\rm IR}=4.5~{\rm GeV}^{-1}$)
of $E_{\rm imp}(r)$ is identified
with $2m_{\rm pole} + V_{\rm QCD}(r)$ up to 
${\cal O}(\alpha_S^3)$, with
the scale $\mu$ fixed to that of the intermediate-distance part
at $r=r_{\rm IR}$, i.e. $\mu = \mu_2(r_{\rm IR})$.
\item
In addition to the ${\cal O}(1/c^3)$ operators considered 
in \cite{Recksiegel:2002za}, 
we include ${\cal O}(1/c^3)$ spin-dependent operators 
\begin{eqnarray}
&&
\delta U_{0}^{(1)} =
\frac{16\pi \alpha_S}{9 m^2 } \, s(s+1)  \, \delta^3 (\vec{r}) \,
\frac{\alpha_S}{\pi}
\left( \frac{23}{12} - \frac{3}{4}\log 2 - \frac{5}{18} n_l \right) , 
\\ &&
\delta U_{0}^{(2)} =
\frac{16\pi \alpha_S}{9 m^2 } \, s(s+1)  \, \frac{\alpha_S}{\pi}
\left[ \frac{21}{4}\, X(r,m) - \frac{\beta_0}{2} \, X(r,\mu) \right]
\label{delU(2)}
,
\\ &&
X(r,\sigma)= \nabla^2 \Bigl[ \frac{1}{4\pi r}
\{ \log (\sigma r) + \gamma_E \} \Bigr] ,
\end{eqnarray}
where $\beta_0 = 11 - \frac{2}{3}n_l$,
and $\gamma_E = 0.5772\dots$ is the Euler constant;
$n_l$ is the number of active quark flavours.\footnote{
For numerical evaluation,
we use the following formula, obtained via integration
by parts and the Schr\"odinger equation with respect to the zeroth-order
Hamiltonian:
$$
\frac{\displaystyle
\int d^3\vec{x}\, |\psi (\vec{x})|^2 \, X(r,\sigma)}
{\displaystyle \int d^3\vec{x}\, |\psi (\vec{x})|^2}
=\frac{\displaystyle
\int_0^\infty dr \, \frac{r}{2\pi} \left[ m \, 
\Bigl\{ V(r)+\frac{l(l+1)}{mr^2}-E \Bigr\} |R_l(r)|^2
+ | R\,'_l(r) |^2 \right] \, \{ \log (\sigma r) + \gamma_E \}
}{\displaystyle \int_0^\infty dr \, r^2 \, |R_l(r)|^2
} ,
$$
where $\psi(\vec{x}) = R_l(r) \, Y_{lm}(\theta ,\phi)$
denotes an unnormalised zeroth-order wave function.
}
The momentum-space representation of these operators can be found
in \cite{BNT}.
These operators 
contribute to the hyperfine splittings at ${\cal O}(\alpha_S^5 m_b)$,
whereas they do not affect the fine splittings at this order.
\item
We also compute the fine splittings and hyperfine splittings of the
charmonium spectrum using the same formalism.
As for the QCD potential, we use the potential constructed
for the bottomonium $E_{\rm imp}(r)$, whose essential part is
given by
$E^{b\bar{b}}_{\rm tot}(r) = V_{\rm QCD}(r) + 2 m_{b,{\rm pole}}
$.\footnote{
The charmonium level splittings computed in this paper
are not affected by an $r$-independent constant to be added to
$E_{\rm imp}(r)$.
In calculating the whole charmonium spectrum, we may
fix the $r$-independent constant
e.g.\ by matching the computed $J/\psi$ mass
to the experimental value.
}
The potential, 
$E_{\rm tot}(r) = V_{\rm QCD}(r) + 2 m_{{\rm pole}}$,
is in principle independent of the quark
masses (apart from an $r$-independent constant).
But it was shown in \cite{Recksiegel:2001,Recksiegel:2002um} that the
use of our specific scale-fixing prescription introduces
a dependence on the quark mass. 
It was shown that (by coincidence) the potential becomes most stable 
for masses of the order of the $b$ quark mass. 
On the other hand, it turns out to be poorly
stable for the charm quark mass.
Furthermore, we confirmed that the potentials corresponding to different
values of quark masses agree with one another
within the estimated theoretical
uncertainties. 
We therefore
use the potential as constructed with the $b$ quark mass both
for bottomonium and charmonium. 
\end{enumerate}
Numerically 
effects of the change (I) are very small, since the fine and hyperfine
splittings are determined predominantly
by the short-distance part of the potential.
(The last digit of every number listed in Tab.~\ref{fine+hyperfine}
varies at most by one.)
Nevertheless, conceptually, $E_{\rm imp}(r)$ is now determined by
perturbative QCD (in a certain prescription) at every $r$.
In particular, we know exactly which higher order corrections
are incorporated in $E_{\rm imp}(r)$ at every $r$.
Therefore, we can estimate effects of the higher order corrections
not included in our calculations, by changing prescriptions or by including
specific types of corrections.
\medbreak

\begin{table}
\begin{center}
\begin{tabular}{|c||c|c|c|c|c|c|c|c|c|}
\hline
\hbox{ \raise-7pt\hbox{Level splitting} } 
& \hbox{ \raise-7pt\hbox{Exp.} } 
& \multicolumn{4}{c|}{Potential model} & 
\multicolumn{2}{c|}{Lattice} & \multicolumn{2}{c|}{Pert.\ QCD based}
\\
\cline{3-10}
 &  & \cite{Eichten:1994gt} 
& \cite{Motyka:1997di} & \cite{EFG} & \cite{fekete} & \cite{CP-PACS} & \cite{LM} &
\cite{PT}  & this work \\
\hline
$\chi_{c1}(1P)- \chi_{c0}(1P)$ & $95$ & $50$ & $81$ & 86 & 72 & 79 & $-$ & $-$ & $56$\\
$\chi_{c2}(1P)- \chi_{c1}(1P)$ & $46$ & $21$ & $50$ & 46 & 49 & 35 & $-$ & $-$ & $43$\\
$J/\Psi - \eta_c(1S)$   & $117$ & $117$ & $117$ & 117 & 117 & 85 & $-$ & $-$ & $88$\\
$\Psi(2S) - \eta_c(2S)$ & $92/ 32$ & $78$ & $72$ & 98 & 92 & 43 & $-$ & $-$ & $38$\\
$\chi_{c}^{\rm cog}(1P) - h_c(1P) $ 
                & $-0.9$ & $0$ & $0$ & 0 & 9 & $1.5$ & $-$ & $-1.4$ & $-0.8$\\
\hline
$\Upsilon(1S) - \eta_b(1S)$ & $(160)$ & $87$ & $57$ & 60 & 45 & $-$ & 51  & $-$ & $44$ \\
$\Upsilon(2S) - \eta_b(2S)$ & $-$     & $44$ & $28$ & 30 & 28 & $-$ & $-$   & $-$ & $21$ \\
$\Upsilon(3S) - \eta_b(3S)$ & $-$     & $41$ & $20$ & 27 & 23 & $-$ & $-$ & $-$ & $12$ \\
$\chi_{b}^{\rm cog}(1P) - h_b(1P)$ & $-$  & $0$ & $0$ & $-1$ & 1 & $-$ & $-$ & $-0.5$ & $-0.4$ \\
$\chi_{b}^{\rm cog}(2P) - h_b(2P)$ & $-$  & $0$ & $0$ & $-1$ & 0 & $-$ & $-$ & $-0.4$ & $-0.2$ \\

\hline
\end{tabular}
\caption{\small
Fine and hyperfine splittings of the charmonium spectrum
and hyperfine splittings of the bottomonium spectrum. 
All values in MeV. 
In models \cite{Eichten:1994gt,Motyka:1997di},
no interaction has been incorporated that produces the 
$^3P_{\rm cog}\!-\!^1\!P_1$ hyperfine splittings.
As for the results of \cite{CP-PACS}, we quote the values from Table 10
of that paper.
In \cite{PT}
the matrix elements of $X(r,m)$ and $X(r,\mu)$ in eq.~(\ref{delU(2)})
were extracted from the experimental values for the fine splittings
instead of computing them from perturbative QCD.
\label{fine+hyperfine}}
\end{center}
\end{table}
In Tab.~\ref{fine+hyperfine}
we list our predictions for
the fine splittings and the hyperfine splittings of the
charmonium spectrum as well as the hyperfine splittings of the
bottomonium spectrum.
Only the states below the threshold for strong decays
($2m_D=3729$~MeV and $2m_B=10558$~MeV) are considered.
$\chi^{\rm cog}_{c,b}$ denotes the centre of gravity of the triplet
$P$-wave states
(the spin-averaged $^3P_J$ mass with the weight factor $2J+1$).
The input parameters of our predictions are
$\alpha_S^{(5)}(M_Z)=0.1181$, 
$m_b^{\overline{\rm MS}}(m_b^{\overline{\rm MS}})=4190$~MeV \cite{bsv2},
$m_c^{\overline{\rm MS}}(m_c^{\overline{\rm MS}})=1243$~MeV \cite{bsv1},
and $\mu=1.5~(3)$~GeV, $n_l=3~(4)$ for charmonium (bottomonium).
For comparison we list the corresponding experimental values,
some model predictions, and
predictions from recent lattice computations.
Our predictions for the bottomonium
fine splittings have already been presented
in \cite{Recksiegel:2002za}.

We see that
the level of agreement of our predictions with the experimental values is
comparable to that of the recent lattice results.\footnote{
It should be noted, however, that qualitatively quite different kinds of
theoretical errors are contained in 
the lattice computations \cite{CP-PACS,LM} and our computations.
For instance, the former
are performed in the quenched approximation, whereas the latter 
include the sea quark effects fully; 
the latter computations are subject to
relativistic corrections, whereas the former computations are 
carried out fully relativistically.
}
On the other hand, generally
the potential models reproduce the experimental values
much better.
This feature would be understandable, 
since the potential models contain much more
input parameters than the lattice or our predictions.
As we will discuss below, our predictions are consistent with the experimental
values with respect to estimated theoretical uncertainties.
The same is true for the lattice results.
On the other hand, generally it is difficult to estimate errors of the
theoretical  predictions of the  potential models (without comparing to
experimental values), as there are no
systematic ways to improve accuracies of their predictions.

As for the splitting $\Psi(2S) - \eta_c(2S)$
in Tab.~\ref{fine+hyperfine}, all the model calculations 
try to reproduce the old experimental value, while the lattice
calculation and our calculation favour the new value.
Note that, even if we take unrealistic values for the input 
parameters of our calculations ($\alpha_S(M_Z)$, 
$m_c^{\overline{\rm MS}}(m_c^{\overline{\rm MS}})$, $\mu$), 
we cannot magnify only the splitting $\Psi(2S) - \eta_c(2S)$
while keeping other splittings almost unchanged.

It has been emphasised \cite{Godfrey}
that the sizes and the signs of the 
$^3P_{\rm cog}\!-\!^1\!P_1$ hyperfine splittings
can distinguish various models,
since there exists wide variation of theoretical predictions 
for this quantity presently.
In particular, \cite{PT} predicted the 
$^3P_{\rm cog}\!-\!^1\!P_1$ hyperfine splittings using a relation between
these splittings and the fine splittings derived from perturbative QCD.
Our predictions are consistent with their values 
within theoretical errors.
The difference is that we predict the 
matrix elements of $X(r,m)$ and $X(r,\mu)$ in eq.~(\ref{delU(2)})
from the input quark masses and the strong coupling constant.
In perturbative QCD, the 
$^3P_{\rm cog}\!-\!^1\!P_1$ hyperfine splittings are generated first
at ${\cal O}(\alpha_S^5 m)$, whereas the fine splittings and the
$S$-state hyperfine splittings are of  ${\cal O}(\alpha_S^4 m)$.
For this reason, and also due to some small overall coefficient, the 
former splittings are predicted to be much smaller
than the latter splittings within perturbative QCD \cite{PT}.
Some potential models do not possess this property,
as discussed in \cite{Godfrey}.

\begin{table}
\begin{center}
\begin{tabular}{|c||c|c|c|c|}
\hline
$\mu$ & 1 GeV & 1.5 GeV & 2 GeV & 3 GeV\\
\hline
$\chi_{c1}(1P)- \chi_{c0}(1P)$ & $61$ & $56$ & $52$ & $47$\\
$\chi_{c2}(1P)- \chi_{c1}(1P)$ & 44   & 43   & 40   & 36  \\
$J/\Psi - \eta_c(1S)$   & $69$ & $88$ & $88$ & $84$\\
$\Psi(2S) - \eta_c(2S)$ & $32$ & $38$ & $38$ & $36$\\
$\chi_{c}^{\rm cog}(1P) - h_c(1P) $ & $-1.3$ & $-0.8$ & $-0.6$ & $-0.4$ \\
\hline 
\end{tabular}\vspace{5mm}
\begin{tabular}{|c||c|c|c|c|c|}
\hline
$\mu$ & 1 GeV & 2 GeV & 3 GeV & 4 GeV & 5 GeV \\
\hline
$\Upsilon(1S) - \eta_b(1S)$ & $ -1$ & $40$ & $44$ & $45$ & $45$\\
$\Upsilon(2S) - \eta_b(2S)$ & $  1$ & $19$ & $21$ & $22$ & $22$\\
$\Upsilon(3S) - \eta_b(3S)$ & $  1$ & $11$ & $12$ & $12$ & $12$\\
$\chi_{b}^{\rm cog}(1P) - h_b(1P)$   
                            & $-1.2$ & $-0.5$ & $-0.4$ & $-0.3$ & $-0.3$\\
$\chi_{b}^{\rm cog}(2P) - h_b(2P)$   
                            & $-0.7$ & $-0.3$ & $-0.2$ & $-0.2$ & $-0.2$\\
\hline
\end{tabular}
\caption{\small
Dependence of our predictions on the scale, all values in MeV. 
         \label{splittingsscaledep}}
\end{center}
\end{table}
The inclusion of the ${\cal O}(1/c^3)$ operators makes our results
fairly scale--independent over a remarkably wide range of scales,
a property that was already emphasised for the bottomonium
fine splittings in
our previous work. We consider this stability against scale variations
an important indication of the reliability of our formalism.
We show the scale dependences of the splittings in 
Tab.~\ref{splittingsscaledep}.

\begin{table}
\hspace{-5mm}
\begin{tabular}{|c||c|c|c|c|c|c||c|}
\hline
Level splitting
& \multicolumn{2}{c|}{(i)$\delta \alpha_S(M_Z)$} 
& (ii)$\frac{1}{2}\Lambda^3r^2$
& (iii)$\mu$-dep. & (iv)NLO & combined &
diff.\ from exp.\ \\
\hline
$\chi_{c1}(1P)- \chi_{c0}(1P)$ & $-14$ & $17$ & 29 & 9 & 15 & 34 & 39 \\
$\chi_{c2}(1P)- \chi_{c1}(1P)$ & $-10$ & $12$ & 21 & 4 & 10 & 24 & 3 \\
$J/\Psi - \eta_c(1S)$   & $-15$ & $18$ & 15 & 19 & $-2$ & 26 & 29 \\
$\Psi(2S) - \eta_c(2S)$ & $-8$ & $9$ & 35 & 6 & 0 & 36 & 54/6 \\
$\chi_{c}^{\rm cog}(1P) - h_c(1P) $ 
                & $0.2$ & $-0.3$ & $-0.4$ & 0.7 & -- & 0.8 & 0.1 \\
\hline
$\Upsilon(1S) - \eta_b(1S)$ & $-6$ & $7$ & 3 & 5 & $-9$ & 11 & (116) \\
$\Upsilon(2S) - \eta_b(2S)$ & $-4$ & $5$ & 6 & 3 & $-4$ & 8 & $-$\\
$\Upsilon(3S) - \eta_b(3S)$ & $-1$ & $3$ & 9 & 1 & $-2$ & 9 & $-$\\
$\chi_{b}^{\rm cog}(1P) - h_b(1P)$ 
                     & $0.1$ & $-0.1$ & $-0.1$ & 0.2 & $-$ & 0.2 & $-$\\
$\chi_{b}^{\rm cog}(2P) - h_b(2P)$ 
                     & $0$ & $-0.1$ & $-0.1$ & 0.1 & $-$ & 0.1 & $-$\\

\hline
\end{tabular}
\caption{\small
Error estimates of our predictions.
See text for explanations on the individual error estimates
(i)--(iv).
As a combined error,
$\sqrt{ ({\rm i})^2 + {\rm Max}\{ {\rm |(ii)|, |(iii)|, |(iv)|} \}^2 }$
is listed.
``diff.\ from exp.'' denotes the difference between our prediction,
given in Tab.~\ref{fine+hyperfine}, and the experimental value.
\label{error}}
\end{table}
We examine several sources of theoretical uncertainties of our predictions
in Tab.~\ref{error}.
Individual errors (i)--(iv) are estimated as follows.
(Unless stated otherwise explicitly,
the same input parameters as in Tab.~\ref{fine+hyperfine} are used.)
\begin{enumerate}
\renewcommand{\labelenumi}{(\roman{enumi})}
\item
Variation of our prediction when $\alpha_S(M_Z)$ is changed from
0.1181 to 0.1161 (left column) or to 0.1201 (right column).
\item
Variation of our prediction when the potential 
$E_{\rm imp}(r)$ is replaced
by $E_{\rm imp}(r) + \frac{1}{2} \Lambda^3 r^2$
($\Lambda=300$~MeV).
According to the analysis \cite{Recksiegel:2001} of $E_{\rm imp}(r)$ 
(see also \cite{Recksiegel:2002um}),
this can be used as an estimate of the error induced by
uncertainties of the potential.
\item
Variation of our prediction when the scale $\mu$ is varied
by a factor of two (between 1--2~GeV for charmonium and 2--4~GeV
for bottomonium).
\item
The size of the next-to-leading order
corrections of our prediction (within our organisation of the perturbative
expansion).
The next-to-leading order corrections to the
$^3P_{\rm cog}\!-\!^1\!P_1$ hyperfine splittings are not yet known.
\end{enumerate} 
Of these, (ii)--(iv) can be regarded as estimates of the sizes of
higher-order corrections.
We have also checked that errors induced by uncertainties of the
input $m_b^{\overline{\rm MS}}(m_b^{\overline{\rm MS}})$ 
and $m_c^{\overline{\rm MS}}(m_c^{\overline{\rm MS}})$ are negligible compared
to the above error estimates, where we varied
$m_b^{\overline{\rm MS}}(m_b^{\overline{\rm MS}})$ 
and $m_c^{\overline{\rm MS}}(m_c^{\overline{\rm MS}})$ 
by 50 and 100~MeV, respectively.

As combined errors, 
$\sqrt{ ({\rm i})^2 + {\rm Max}\{ {\rm |(ii)|, |(iii)|, |(iv)|} \}^2 }$
are listed.
Comparing them with the differences of our predictions from the
experimental values, for charmonium, we find a reasonable agreement
of our predictions and the experimental values with
respect to the estimated errors.
Furthermore, 
the values of the combined errors are consistent with the error estimates,
which can be obtained following the logic employed for estimating theoretical
uncertainties of the bottomonium fine splittings
in \cite{Recksiegel:2002za}:
order 
$\Lambda_{\rm QCD}^3/m_c^2 \simeq 10$--50~MeV for the charmonium
fine and hyperfine splittings and order 
$\Lambda_{\rm QCD}^3/m_b^2 \simeq 1$--10~MeV for the bottomonium
hyperfine splittings,
except for the $^3P_{\rm cog}\!-\!^1\!P_1$ hyperfine splittings.
\medbreak

To conclude,
we find that our formalism gives predictions for
the charmonium fine and hyperfine splittings with moderate
uncertainties that are
comparable to the recent lattice computations.
It provides yet another evidence that
perturbative QCD is compatible with the experimental data within
estimated perturbative uncertainties
\cite{bsv1,bsv2,Recksiegel:2002za}.
Uncertainties of our predictions for the bottomonium hyperfine splittings
are much smaller than those for the charmonium.

The input parameters of our predictions are 
the quark $\overline{\rm MS}$ masses and the strong coupling constant,
apart from the scale that is varied within a reasonable range as part of
the error analysis.
In addition, our predictions are dependent on the
specific prescriptions to define $E_{\rm imp}(r)$ or on the specific
organisation of our perturbative series.
Nonetheless, our predictions together with the estimated
uncertainties lie within the frame of perturbative QCD, 
hence, our formalism can be used for testing the validity of perturbative QCD
in comparison with the experimental data.
In this sense, the freedom we have in performing an analysis based on
perturbative QCD is rather restricted:
either we may find a more sophisticated prescription and
reduce theoretical errors, or choose a poorer prescription and increase 
errors, other than changing the values
of the aforementioned input parameters;
we cannot incorporate any contributions which lie outside of 
perturbative QCD.
This is in contrast with the conventional phenomenological models, 
which typically have much more input parameters; they can
be adjusted so as to reproduce experimental data, without affecting
uncertainties of theoretical predictions.
Thus, we consider it quite non-trivial
that the experimental data are reproduced with respect to moderate
theoretical uncertainties, even though there appears to be some 
arbitrariness in defining our formalism.

Although our formalism predicts the complete spectrum, 
in this article we concentrated on the level splittings
that have smaller theoretical uncertainties than the levels themselves.
We have previously shown that our formalism also reproduces the overall
bottomonium spectrum (as far as it is measured) well \cite{Recksiegel:2002za}. 
The same is true 
for charmonium, although theoretical uncertainties 
are larger in this case \cite{fullpaper}.

\section*{Acknowledgements}
The authors wish to thank S.~Godfrey for valuable suggestions.
Y.S.\ is grateful to T.~Kaneko for information on 
recent lattice results.

\section*{Note added on Nov.\ 20th, 2003:}
When the manuscript for this paper was prepared, only the
(inconsistent) Crystal Ball \cite{Edwards:1981mq} 
and Belle \cite{Choi:2002na} measurements
for the $\eta_c(2S)$ mass were available, so both of them were
quoted in Table 1. Very recently, several more new measurements
have appeared that support the Belle value: BaBar measures
$3630.8 \pm 3.4 \pm 1.0$ MeV \cite{proof1} (an earlier analysis resulted in
$3632.2 \pm 5.0 \pm 1.8$ MeV \cite{proof2}), CLEO II gives
$3642.7 \pm 4.1 \pm 4.0$ MeV (CLEO III prelim.:
$3642.5 \pm 3.6 \pm ?$ MeV) \cite{proof3} and a different Belle
analysis yields $3630 \pm 8$ MeV \cite{proof4}.
The $\psi(2S)-\eta_c(2S)$ splitting calculated from a na\"ive
average of the central values of all measurements except Crystal
Ball is 47 MeV, in excellent agreement with our results.

\end{document}